\documentclass{amsart}

\usepackage{amsmath,amsthm,amssymb,bm}
\usepackage{hyperref}

\usepackage{enumerate}

\numberwithin{equation}{section}

\newcommand{\ndash}{\nobreakdash-\hspace{0pt}}
\newcommand{\Ndash}{\nobreakdash--}

\newcommand{\ii}{{\mathrm{i}}}
\newcommand{\dd}{{\mathrm{d}}}

\newcommand{\barr}[1]{\overline{{#1}}}

\newcommand{\T}{{\mathbb T}}

\newcommand{\be}{\begin{equation}}
\newcommand{\ee}{\end{equation}}

\newcommand{\calL}{{\mathcal{L}}}

\newcommand{\calF}{{\mathcal{F}}}
\newcommand{\calEL}{\mathcal{EL}}

\newcommand{\EE}{\mathrm{e}}

\DeclareMathOperator{\Div}{div}

\DeclareMathOperator{\EL}{EL}
\DeclareMathOperator{\graph}{graph}

\newcommand{\ev}{\mathrm{ev}}

\newtheorem{Thm}{Theorem}[section]

\newtheorem*{Thm*}{Theorem}
\newtheorem*{Lem*}{Lemma}

\theoremstyle{remark}
\newtheorem{Rem}[Thm]{Remark}
\newtheorem*{Ack}{Acknowledgment}

\newtheorem*{Rem*}{Remark}

\theoremstyle{definition}

\newtheorem{Def}[Thm]{Definition}
\newtheorem{Exa}[Thm]{Example}

\newcommand{\bbR}{{\mathbb{R}}}

\newcommand{\bbZ}{{\mathbb{Z}}}
\newcommand{\bbC}{{\mathbb{C}}}

\newcommand{\de}{\partial}

\newcommand{\bx}{\mathbf{x}}
\newcommand{\bq}{\mathbf{q}}
\newcommand{\bv}{\mathbf{v}}

\newcommand{\calA}{\mathcal{A}}
\newcommand{\calB}{\mathcal{B}}
\newcommand{\calH}{\mathcal{H}}
\newcommand{\calS}{\mathcal{S}}
\newcommand{\calC}{\mathcal{C}}

\newcommand{\calD}{\mathcal{D}}

\newcommand{\calE}{\mathcal{E}}

\def\gpd{\,\lower1pt\hbox{$\longrightarrow$}\hskip-.24in\raise2pt
               \hbox{$\longrightarrow$}\,}

\let\Tilde=\widetilde

\let\Hat=\widehat

\newcommand\qq{}
\newcommand\cmp[1]{{\qq Commun.\ Math.\ Phys.\ \bf #1}}

\newcommand\pl[1]{{\qq Phys.\ Lett.\ \bf #1}}

\newcommand\ijmp[1]{{\qq Int.\ J. Mod.\ Phys.\ \bf #1}}

\newcommand\jdg[1]{{\qq J.\ Diff.\ Geom.\ \bf #1}}

\newcommand\inm[1]{{\qq Invent.\ Math.\ \bf #1}}

\begin{document}

\title{Classical and Quantum Lagrangian Field Theories with Boundary}


\author[A.~S.~Cattaneo]{Alberto~S.~Cattaneo}
\address{Institut f\"ur Mathematik, Universit\"at Z\"urich,
Winterthurerstrasse 190, CH-8057 Z\"urich, Switzerland}
\email{alberto.cattaneo@math.uzh.ch}

\author[P. Mnev]{Pavel Mnev}
\address{Institut f\"ur Mathematik, Universit\"at Z\"urich,
Winterthurerstrasse 190, CH-8057 Z\"urich, Switzerland}
\email{pmnev@pdmi.ras.ru}

\author[N. Reshetikhin]{Nicolai Reshetikhin}
\address{Department of Mathematics,
University of California, Berkeley,
CA 94720, USA}
\email{reshetik@math.berkeley.edu}

\thanks{A.~S.~C. acknowledges partial support of SNF Grant No.~200020-131813/1.
P.~M. acknowledges partial support of RFBR Grants
Nos.~11-01-00570-a and
11-01-12037-ofi-m-2011
and of SNF Grant No.~200021-137595.}

\begin{abstract}
This note gives an introduction to Lagrangian field theories in the
presence of boundaries.
After an overview of the classical aspects, the cohomological formalisms
to resolve singularities in the bulk and in the boundary theories (the BV
and the BFV formalisms, respectively) are recalled.  \
One of the goals here
(and in \cite{CMR}) is to show how the latter two formalisms can be put
together in a consistent way, also in view of perturbative quantization.
\end{abstract}

\maketitle
\tableofcontents

\section{Introduction}\label{intro}
Lagrangian field theories are used to describe physical models. Their quantization is somehow expected to satisfy Segal's \cite{S} axioms (which generalize in higher dimensions our understanding of quantum mechanics). Roughly speaking they say that a $d$\ndash dimensional quantum field theory should provide
a functor from the category of $d$\ndash dimensional cobordisms (with possible extra structures) to the category of vector spaces and linear maps.
This program has been completed in conformal field theories. Atiyah proposed a stricter version of these axioms in the case of topological field theories,
which was first realized by the Reshetikhin--Turaev invariant \cite{RT} which may be thought of as some nonperturbative quantization of Chern--Simons
theory \cite{W,FK}. Another example is implicitly present in an old work
by Migdal \cite{Mig}.

It would be interesting to understand how this picture emerges from the perturbative functional integral quantization of Lagrangian fields theories.
An important  application would
be the construction of perturbative quantum field theories on manifolds out of topologically simple or geometrically small pieces, where the computations might be more tractable.
Even in the case of topological field theories this would produce new insight (and might possibly lead to a full understanding of the relation between the
perturbative expansion of Chern--Simons theory and the asymptotics of the Reshetikhin--Turaev invariant).

The first step in developing this program, performed in \cite{CMR}, consists in developing the analogous pictures in the classical formalism and in the BV formalism \cite{BV} which is the main tool for the perturbative quantization of theories with symmetries. In principle, this already  yields the possibility of constructing moduli spaces of solutions to variational problems out of computations on topologically simple or geometrically small pieces.

This note reviews some results of \cite{CMR}, see Section~\ref{S:BV},
with a didactical introduction through classical Lagrangian field theory given in Sections~\ref{S:LI} and \ref{S:LII},
and with the BFV formalism \cite{BFV}, being outlined in Section~ \ref{S:BFV}. Notice that Section~\ref{S:BV} is
self contained, so the hasty reader who does not need  a motivation or an introduction might well jump directly there.

As a final remark, notice that in this paper every manifold is assumed to be compact, though possibly with boundary.





\begin{Ack}
We thank F. Bonechi, H. Bursztyn, A. Cabrera, K. Costello, C. De Lellis, G. Felder, V. Fock and E. Getzler for useful discussions.
We especially thank J.Stasheff for helpful comments on a first draft.
A.S.C. thanks University of Florence, IMPA and Northwestern University for hospitality. 
\end{Ack}

\section{Lagrangian field theory I: Overview}\label{S:LI}
We start reviewing classical Lagrangian mechanics. This is usually defined by specifying a Lagrangian function $L$ on the tangent bundle $TN$ of
some manifold $N$. The action $S_{[t_0,t_1]}$ corresponding to an interval $[t_0,t_1]$ is a function on the path space
$N^{[t_0,t_1]}$ defined by
\begin{equation}\label{e:S}
S_{[t_0,t_1]}[x]=\int_{t_0}^{t_1} L(\dot x(t),x(t))\;dt.
\end{equation}
The Euler--Lagrange (EL) equations describe the critical points of the action. As this requires integration by parts, one usually puts appropriate boundary conditions for the boundary terms to vanish, e.g., one fixes the initial and final values $x(t_0)$ and $x(t_1)$.
In the sequel we will want to avoid this.

\subsection{Symplectic formulation}
The typical example is Newtonian mechanics on a Riemannian manifold. In this
case $N$ is a Riemannian manifold and  the Lagrangian is $L(q,v)=\frac12 m ||v||^2 - V(q)$, where $||\ ||$ is the norm induced by a metric on $N$, $V$ is a function on $N$ and $m$ is a parameter (usually assumed to be strictly positive). The EL equations in this case are the Newton equations with force given by $-\nabla V$. They admit, locally, a unique solution if initial position and velocity are specified. We denote by $C=TN$ the space of such initial conditions. Here $C$ stands here for Cauchy, but we will see later it can also stand for coisotropic. Notice the peculiar coincidence that $C$ is the same as the space on which $L$ is defined. This will not be the case in further examples.

One usually reformulates the problem in symplectic terms using the Legendre mapping $\phi_L\colon TN\to T^*N$, $(v,q)\mapsto (p(v,q),q)$, with $p_i=\frac{\de L}{\de v^i}$. The Newton equations of motions now become first order and their solution yields the symplectic flow $\Phi_{H_L}$ with respect to the canonical symplectic form $\omega_\text{can}$ on $T^*M$ and Hamiltonian function $H_L$ given by the Legendre transform of $L$: $H_L(p,q)=p_i v^i(p,q) - L(v(p,q),q)$.
Here we have used the inverse of the Legendre mapping $\phi_L^{-1}\colon T^*N\to TN$,
$(p,q)\mapsto (v(p,q),q)$, where $v(p,q)$ is the inverse to $p(v,q)$. Since in the following we will also be interested in degenerate Lagrangians, for which the Legendre mapping is not a local diffeomorphisms, we now recall how to reformulate things without going to the cotangent bundle. The first simple fact is that
\begin{equation}\label{e:alphaL}
\alpha = \frac{\de L}{\de v^i}dq^i
\end{equation}
is a well-defined one\ndash form on $C=TN$ (it would be more precise to write $\alpha_L$ instead of $\alpha$ to stress the dependency on $L$, but we regard $L$ as given for the whole discussion).
Moreover, $\omega:=\dd\alpha$ is  non degenerate precisely when the Lagrangian is regular, i.e. when the Legendre mapping is a local diffeomorphism. In this case, one can also easily show that $\omega=\phi_L^*\omega_\text{can}$. We can now formulate the Hamiltonian evolution directly on $C$. For later considerations, however, it is better to consider the graph of the Hamiltonian flow instead of the flow itself. Borrowing notations from the cotangent bundle, we then consider
$L_{[t_0,t_1]}=(\phi_L^{-1}\times \phi_L^{-1})(\graph(\Phi_{H_L}|_{t_0}^{t_1}))$. As a Hamiltonian flow is a symplectomorphism and as the graph of a symplectomorphism is a Lagrangian submanifold  of the Cartesian product with reversed sign of the symplectic form on the first factor, we have that $L_{[t_0,t_1]}$ is Lagrangian submanifold in $\bar C\times C$ . The fact that it comes from the graph of a flow yields the property
\begin{equation}\label{e:Ltt}
L_{[t_1,t_2]}\circ L_{[t_0,t_1]} = L_{[t_0,t_2]},\quad \lim_{t_1\to t_0} L_{[t_0,t_1]}=\mbox{ graph of the Id map }.
\end{equation}
The limit has to be understood by putting an appropriate topology on the space of submanifolds of $C$.

The crucial point now is that $L_{[t_0,t_1]}$ may be defined directly without making reference to the Hamiltonian flow. Let
\[
\pi_{[t_0,t_1]}\colon\begin{array}[t]{ccc}
N^{[t_0,t_1]} &\to & C\times C\\
\{x(t)\} &\mapsto &((\dot x(t_0),x(t_0)),(\dot x(t_1),x(t_1)))
\end{array}
\]
and let $EL_{[t_0,t_1]}\subset N^{[t_0,t_1]}$ be the space of solutions to the EL equations. Then we simply have
\begin{equation}\label{e:LpiEL}
L_{[t_0,t_1]}=\pi_{[t_0,t_1]}(EL_{[t_0,t_1]}).
\end{equation}
This is the fundamental equation we are going to use. Notice that \eqref{e:Ltt} immediately follow from this new definition (the limiting property follows from the fact that $C$ is the space of initial conditions that guarantee local existence and uniqueness).

To insure that $L_{[t_0,t_1]}$ is Lagrangian  we have to make some further observations. First we observe that, if we do not impose boundary conditions, the variational calculus yields
\begin{equation}\label{e:deltaS1d}
\delta S_{[t_0,t_1]} = \EL_{[t_0,t_1]} + \pi_{[t_0,t_1]}^*\alpha,
\end{equation}
where $\EL_{[t_0,t_1]}$
is the term containing the EL equations that we reinterpret as a one\ndash form on $N^{[t_0,t_1]}$. Notice that we also interpret the variation symbol $\delta$ as the
de~Rham differential on $\Omega^\bullet(N^{[t_0,t_1]})$. The appearance of the same $\alpha$ here and in \eqref{e:alphaL} is crucial. The equation yields in fact, after differentiation,
$\pi_{[t_0,t_1]}^*\omega =-\delta \EL_{[t_0,t_1]}$, which in turn implies that the restriction of $\omega$ to $L_{[t_0,t_1]}$ vanishes (observe here that the space of solutions to the EL
equations on $M^{[t_0,t_1]}$ is the zero locus of
$\EL_{[t_0,t_1]}$ and that
$\pi_{[t_0,t_1]}$ is a surjective submersion). This amounts to saying that $L_{[t_0,t_1]}$ is isotropic.

In addition we know that a unique solution is given, locally, once we specify initial conditions in $TN$ or, equivalently, if we specify initial and final position for a short enough time interval $[t_0,t_1]$. Hence, if we fix initial and final positions $Q_0$ and $Q_1$ and let $L_{Q_0,Q_1}:=\{(v,q,v',q')\in C\times C | q=Q_0,\ q'=Q_1\}$, we get that, for $t_1$ sufficiently close to $t_0$,
$L_{Q_0,Q_1}\cap L_{[t_0,t_1]}$ consists of one point; by dimension counting, assuming the intersection
 is transversal, this implies that $L_{[t_0,t_1]}$ has half dimension than $C$ and so is Lagrangian\footnote{Here is another argument. For a short interval,
the initial and final positions specify a
unique solution. Hence initial and final positions determine initial and final
velocities, which implies the L is a graph, hence Lagrangian.}. Finally, if $t_0$ and $t_1$ are not close, we can decompose the interval into short ones on which we can use the previous argument and recover $L_{[t_0,t_1]}$ as the composition of the canonical relations corresponding to the subintervals. Hence $L_{[t_0,t_1]}$ is also a canonical relation.

After understanding this, we can also think of more general boundary conditions by replacing $L_{Q_0,Q_1}$ with another submanifold $L$ of $\barr C\times C$
on which $\alpha$ vanishes. The latter condition ensures that the variational problem has no boundary contributions. It also implies that $L$ is isotropic.
In order to have, generically, intersection points of $L$ with $L_{[t_0,t_1]}$, one has to require $L$ to have maximal dimension, and hence to be Lagrangian\footnote{ If $L$ is Lagrangian but $\alpha$ does not vanish on it, we can modify the action by adding boundary terms such that the modified one-form $\alpha$ is vanishing on $L$}. Now, the intersection
$L\cap L_{[t_0,t_1]}$ can be considered as the space of solutions to the EL equations. Notice however that this intersection might as well be empty or contain (infinitely) many points, though generically, it will be a discrete
set.

\subsection{A degenerate example: geodesics on the Euclidean plane}\label{s:geod}
We now consider the non-regular Lagrangian $L(v,q):=||v||$, where $||\ ||$ is the Euclidean norm and $v,q\in T\bbR^2$. The action is still given by \eqref{e:S} which we now define only on the space $N_0^{[t_0,t_1]}$ of
 immersed paths (i.e., we impose the condition $\dot\gamma(t)\not=0$ $\forall t\in[t_0,t_1]$). The EL equations have as solutions parameterized segments of straight lines in $\bbR^2$. By analogy with Newtonian mechanics the Cauchy data include initial position and initial velocity, but these data do not give uniqueness: they define one end of the segment and
 its slope but do not define a parametrization of the
 segment uniquely. Nevertheless, let us consider what will happen if we take the ``wrong" space of Cauchy data $C=T\bbR^2\setminus\{$zero section$\}$. We defer to
 Remark~\ref{r:CauchyE} the discussion of the ``true" space of Cauchy data.

Using polar coordinates for the velocities, we can write $C=\bbR^2\times S^1\times\bbR_{>0}$ with coordinates the position $\bq$, the normalized velocity $\bv$ (i.e. $||\bv ||=1$) and the speed $\rho$.
Then \eqref{e:alphaL} yields $\alpha=\bv\cdot\dd\bq$, so that $\omega=\dd\bv\cdot\dd\bq$ is clearly degenerate (as it does not have a $\dd\rho$ component).
On the other hand, \eqref{e:LpiEL} (with $\pi$ the evaluation of  position and velocity at the initial and at the final point)
yields
\[
L_{[t_0,t_1]}=\{(\bq_0,\bv,\rho_0),(\bq_1,\bv,\rho_1))| (\bq_1-\bq_0)\, \mbox{ parallel to }\,\bv\}
\]
which is clearly not a graph but can easily be checked to be Lagrangian in $\barr C\times C$ (see Appendix~\ref{a:rec} for the definition of Lagrangianity in the case of a degenerate two\ndash form).

In this example one can can easily get rid of the degeneracy of $\omega$ by taking the quotient by its kernel which is the span of the vector fields
$\bv\cdot\frac\de{\de\bq}$ and $\frac\de{\de\rho}$ (geometrically, these vector fields represent the space directions parallel to the velocity and the rescalings of the velocity). The quotient turns out to be
$TS^1$ with symplectic structure given by pullback of the canonical one on $T^*S^1$ by the induced metric on $S^1$. Notice that the base $S^1$ here is  the space of normalized velocities, whereas the tangent fiber can be thought of as the space direction orthogonal to the given velocity. One can also project $L_{[t_0,t_1]}$ down to the quotient $TS^1\times TS^1$ . The result is just the graph of the identity map on $TS^1$. This is a consequence of the fact that the action is invariant under reparametrization. Notice that this is an example of topological theory: the action does not depend of the metric on $[t_0,t_1]$.

In this example we passed to the quotient space which appeared to be smooth. However that in general reduction may produce very singular quotients, so passing directly to the reduced space had better be avoided. Instead, as we will see, it is better to use the BV-BFV approach. 

\begin{Rem}\label{r:redconn}
Notice that the one\ndash form $\alpha=\bv\cdot d\bq$ is not horizontal with respect to the kernel of $\omega$, so it cannot be reduced to $TS^1$. On the other hand, we may regard $\alpha$ as a connection one\ndash form on the trivial line bundle on $C$. We can then reduce this line bundle to a line bundle over $TS^1$ and reduce $\alpha$ as a connection.

Also notice that evaluating the action on a solution yields a well-defined function $S_{HJ}$ (the Hamilton--Jacobi action) on $L_{[t_0,t_1]}$ which is just the length of the path.
Again, $S_{HJ}$ cannot be reduced to a well-defined function on $\overline{TS^1}\times TS^1$, but $\exp\frac\ii\hbar S_{HJ}$ can be reduced to a section of the reduced line bundle.
\end{Rem}

\begin{Rem}
If one considers the same example but now with the Minkowski metric, also in higher dimensions, and considers only timelike velocities (i.e., with $\dd s^2=\dd t^2-\dd \bx^2$, one assumes $||v||>0$), the reduction yields $TH$, where $H$ is the upper hyperboloid $v_0^2-\bv^2=1$, and the symplectic structure is obtained by pullback of the canonical one on the cotangent bundle by the hyperbolic metric on $H$ (which is induced by the Minkowski metric). The reduced Lagrangian is again just the graph of the identity.
\end{Rem}

\begin{Rem}\label{r:CauchyE}
We mentioned above that $T\bbR^2\setminus\{$zero section$\}$ is not the true space of initial conditions because giving initial position and velocity does not select a unique parametrized segment. In order to obtain uniqueness in a formal neighborhood
of the initial point we can enlarge $C$ by setting $\Tilde C=T\bbR^2\setminus\{$zero section$\}\times\bbR^\infty$, where
the coordinates on $\bbR^\infty$ are all higher derivatives of the path. In other words here we work with the space of jets. Define the one form $\tilde{\alpha}$ as the pull-back of $\alpha$ with respect to the natural projection $\tilde{C}\to C$. It is clear the $\bbR^\infty$ factor is in the kernel of $\tilde{\alpha}$ and of $\tilde{\omega}=\omega$. This is the reason why this factor can be completely neglected just as we  neglected the space $\bbR>0$ of speeds in the discussion above.
The reduction of $\Tilde C$ is the same as that of $C$.
Boundary values \eqref{e:LpiEL} of solutions to the Euler-Lagrange equations define a  Lagrangian subspace in $\barr{\Tilde C}\times {\Tilde C})$. Its reduction is again the graph of the identity.
\end{Rem}

\subsection{Example: Free scalar field theory}\label{s:free}
We now describe an example of a field theory in dimension $d$. The space time in such theory is a
Riemannian $d$\ndash manifold $(M,g_M)$. The space of fields is $\bbR^M$ (functions on $M$) and the action is
\[
S_{(M,g_M)}[\phi]=\frac{1}{2}\int_M g_M^{\mu\nu}\de_\mu\phi\,\de_\mu\phi\; \dd vol_{g_M}=\frac{1}{2}\int_M (d\phi, d\phi) \dd vol_{g_M}
\]
where $\phi\in\bbR^M$. Solutions to the EL equations are harmonic functions on $M$.

In order to understand the boundary structure for an arbitrary
space time manifold $M$ consider first a thin neighborhood
of its boundary $\Sigma= \partial M$. That is consider a short cylinder $\Sigma\times[0,\epsilon]$ where $(\Sigma,g_\Sigma)$ is a $(d-1)$\ndash dimensional Riemannian manifold and $g_M=g_\Sigma+\dd s^2$, $s\in[0,\epsilon]$.
A unique solution to the EL equation is obtained if one specifies the values of $\phi$ and of its normal derivative on $\Sigma\times\{0\}$. This gives the natural space of Cauchy data associated to $\Sigma$,  $C_{\Sigma}=\bbR^{\Sigma}\times\bbR^{\Sigma}$. Similarly to \eqref{e:alphaL} the boundary term in the variation of $S_{\Sigma\times[0,\epsilon]}$ gives
the one\ndash form
\[
\alpha_{(\Sigma,g_\Sigma)}=\int_{\Sigma} \chi\,\delta \phi\;\dd vol_{g_\Sigma}
\]
with $(\chi,\phi)\in C_{\Sigma}$. Here $\chi$ should be thought of as the restriction  to $\Sigma\times\{0\}$ of the normal derivative of the bulk field $\phi$. Notice that $\omega_{(\Sigma,g_\Sigma)}=\delta \alpha_{(\Sigma,g_\Sigma)}$ is (weakly) nondegenerate.

For a general Riemannian $d$\ndash manifold $M$ with boundary $\de M$,  we have the surjective submersion
$\pi_M\colon \bbR^M\to C_{\de M}$
obtained by evaluating the field $\phi$ and its normal derivative on the boundary.
Formula
\eqref{e:deltaS1d} still holds (with a little change in notation):
\[
\delta S_{(M,g_M)} = \EL_{(M,g_M)} + \pi_{M}^*\alpha_{(\de M,g_M|_{\partial M})}.
\]
Moreover, in the spirit of \eqref{e:LpiEL}, define $L_{(M,g_M)}\subset C_{\de M}$ as $\pi_M(EL_{(M,g_M)})$.

It is easy to see that $L_{(M,g_M)}$ is Lagrangian. Indeed,
the Dirichlet problem for $\phi$ has unique solution on $M$.
Thus, if $\phi\in EL_M$, its boundary values define the
the normal derivative of $\phi$ at the boundary.
This map from the Dirichlet data to Neumann data is
known as the Dirichlet-to-Neumann mapping. Thus,
the submanifold $L_M\in C_{\partial M}$ is the graph of the Dirichlet-to-Neumann mapping $\bbR^{\partial M}\to \bbR^{\partial M}$.





Notice that one may distinguish the connected components of $\de M$ into incoming and outgoing: $\de M=\de_\text{in} M\sqcup\de_\text{out} M$.
Denoting $\de_\text{in} M$ with opposite orientation by $\de_\text{in} M^\text{opp}$, we may then  view
$L_{(M,g_M)}\subset \barr{C_{\de_\text{in} M^\text{opp}}}\times C_{\de_\text{out} M}$
as a canonical relation from $C_{\de_\text{in} M^\text{opp}}$ to $ C_{\de_\text{out} M}$.

\subsection{Conclusions}
{}From these examples we see that the Hamiltonian framework for
non-regular Lagrangians has to be replaced by its weaker version. However, certain important patterns remain. We have seen in these
examples that that $i)$ in all cases we were able to derive
a, possibly degenerate,  two\ndash form on the space of initial conditions associated to the boundary of the space time and $ii)$ we were able to assign to the bulk of the space time a Lagrangian/isotropic  submanifold (not necessarily a graph)
in such spaces. In the next Section we will see that this a quite general fact.

\section{Lagrangian field theory II (after V.~Fock)}\label{S:LII}
In an unpublished account \cite{F}, V.~Fock has considered the general structure of Lagrangian field theories on manifolds with boundary.
We give here our recapitulation of this account. (To different levels of generality, this structure has been rediscovered many times, see e.g.\  \cite{Gawedzki,Schwarz} and references therein.)

Notice that this Section is rather about a philosophical account leading to a concrete construction than a precise mathematical formulation. For simplicity, we also assume that all the "spaces" occurring in the following are actually (possibly infinite dimensional) manifolds.

A Lagrangian field theory is specified by
\begin{enumerate}
\item  fixing the dimension $d$ of the source manifolds;
\item fixing a class of $d$\ndash manifolds, possibly with extra structure, such as a metric in the example of subsection~\ref{s:free};
\item associating a space of fields $F_M$ (functions, maps to a fixed target manifold, sections of bundles, connections,\dots)  to every $d$\ndash manifold $M$ in the class;
\item defining a density, the Lagrangian $L$, of the fields and finitely many of their derivatives.\footnote{For a precise definition, see, e.g., \cite{C}}
\end{enumerate}
The action functional associated to a manifold $M$, as a function on $F_M$, is then given by $S_M=\int_M L$. The variation of the action, neglecting the boundary terms, yields
the EL equations. We denote by $EL_M\subset F_M$ the space of solutions to the EL equations on a given manifold $M$.

Let now $\Sigma$ be a $(d-1)$\ndash manifold. We extend it to a $d$\ndash manifold $M:=\Sigma\times[0,\epsilon]$ (taking care of the possible additional structure).
The variational calculus on this particular $M$ produces two new pieces of data:
\begin{enumerate}
\item The space $C_\Sigma$ of Cauchy data consisting on the information on the fields (and their derivatives) that one has to specify on $\Sigma$ so that there is
a unique solution to EL equations on $\Sigma\times[0,\epsilon]$ for $\epsilon$ small enough (possibly, one might have to work with a formal neighborhood of $0$ like in the example
of subsection~\ref{s:geod}).
\item A one\ndash form $\alpha_\Sigma$ on $C_\Sigma$ arising from the $\Sigma\times\{0\}$\ndash boundary contribution to the variation of $S_{\Sigma\times[0,\epsilon]}$.
\end{enumerate}
One can see that the Lagrangian is regular if{f} $\omega_\Sigma:=\dd\alpha_\Sigma$ is non degenerate.

Using these data, one can further develop the induced structure. Namely, for every $M$ in the class we now have a surjective submersion
\[
\pi_M\colon F_M\to C_{\de M}
\]
and the variation of the action leads to the fundamental equation
\begin{equation}\label{e:fundamentalFock}
\delta S_M = \EL_M + \pi_M^*\alpha_{\de M}.
\end{equation}
Again we define $L_M:=\pi_M(EL_M)\subset C_{\de M}$. It follows from \eqref{e:fundamentalFock} that $L_M$ is isotropic (i.e., the restriction of $\omega_{\de M}$ vanishes).
In most examples $L_M$ is actually Lagrangian.

\begin{Rem}
{}From now on for simplicity we are going to assume that all $F_M$ and $C_\Sigma$ can be given a manifold structure and that $EL_M$ and $L_M$ are smooth submanifolds and, apart from the following counterexample, we are always going to assume that $L_M$ is Lagrangian.
\end{Rem}

\begin{Exa}[$L_M$ is not Lagrangian]\label{e:nonlag}
Consider a one\ndash dimensional example with target $\bbR^2$, space of fields $F_{[t_0,t_1]}=(\bbR^2)^{[t_0,t_1]}$, with Lagrangian $L\in C^\infty(T\bbR^2)$
given by $L(v_x,v_y,x,y)=\frac12 y\,v_x^2$ and with the
action $S_{[t_0,t_1]}=\int_{t_0}^{t_1} \frac 12 y(t)\,\dot x(t)^2\;\dd t$. This example is related to the one given in \cite{HT}(section 1.2.2)  where it is used to disprove a related conjecture by Dirac. Note also that this is a 1-dimensional version of the Polyakov string action.

The EL equations are $\dot x^2=0$ and $\frac\dd{\dd t}(y\,\dot x)=0$, the latter being trivially implied from the first. The $x$\ndash component of a solution is then completely determined by its initial value, whereas the $y$\ndash component is completely free. To get formal uniqueness of solutions to the EL equations  define $C=\bbR\times\bbR^\infty$. Here the second factor contains the information about $y$ and all its derivatives at the initial time. The variation of the action is $\int_{t_0}^{t_1} (\frac{1}{2}
\delta y \dot x^2+ y\,\dot x\,\delta\dot x) \;\dd t$. However the boundary term here is absent because we have to assume $\dot x=0$ on the boundary in order to have a solution and therefore $\alpha=0$. The projection $\pi_{[t_0,t_1]}\colon F_{[t_0,t_1]}\to C\times C$ is then simply given by
\[
\pi(x(\cdot),y(\cdot))= (x(t_0),y(t_0),\dot y(t_0),\ddot y(t_0),\dots,x(t_1),y(t_1),\dot y(t_1),\ddot y(t_1),\dots)
\]
so that $L_{[t_0,t_1]}=\{(x,y_0,y_1,y_2,\dots,x,\tilde y_0,\tilde y_1,\tilde y_2,\dots), x,y_i,\tilde y_i\in\bbR\ \forall i \}$.
Now $L_{[t_0,t_1]}$ is obviously isotropic since $\omega=0$.  On the other hand, since $\omega=0$, $L_{[t_0,t_1]}^\perp=C\neq L$. Hence $L_{[t_0,t_1]}$ is not Lagrangian.
\end{Exa}
As in subsection~\ref{s:free}, we can decide to split the boundary of a given $d$\ndash manifold $M$ into incoming and outgoing boundary components,
$\de M=\de M_\text{in}\sqcup\de M_\text{out}$, and regard $L_M$ as a canonical relation from $C_{\de M_\text{in}^\text{opp}}$ to $ C_{\de M_\text{out}}$, which we will call
the \textsf{evolution relation} since it generalizes the evolution flow.

Suppose we cut a manifold $M$ along a submanifold $\Sigma$ into two manifolds $M_1$ and $M_2$ in such a way that $\de_\text{in} M\subset\de M_1$ and
$\de_\text{out}M\subset\de M_2$. Then we set $\de_\text{in}M_1=\de_\text{in}M$, $\de_\text{out}M_1=\Sigma$, $\de_\text{in}M_2=\Sigma^\text{opp}$ and
$\de_\text{out}M_2=\de_\text{out}M$. We then have
\[
L_M=L_{M_2} \circ L_{M_1}
\]
since a solution on $M$ corresponds to solutions on $M_1$ and $M_2$ that match on $\Sigma$. This composition of canonical relations replaces the usual composition of flows.

In particular, in the case of a cylinder $\Sigma\times{[t_0,t_1]}$ we have
\[
L_{\Sigma\times[t_0,t_1]}=L_{\Sigma\times[t,t_1]}\circ L_{\Sigma\times[t_0,t]}
\]
for all $t\in(t_0,t_1)$. On the other hand, in general we cannot expect $L_\Sigma:=\lim_{t_1\to t_0} L_{\Sigma\times[t_0,t_1]}$ to be the graph of the identity on $C_\Sigma$
as this will happen only in regular theories. We call gauge theories those for which $L_\Sigma$, which can be seen to be an equivalence relation, is not the graph of the identity.

\begin{Rem}[Evolution correspondences]
Recall that the Cauchy space $C_\Sigma$ determines uniqueness only locally (or just formally, like in example~\ref{e:nonlag} or in remark~\ref{r:CauchyE}) on cylinders.
Therefore, we may miss some important information by looking only at $L_M$. The information about nonuniqueness is contained in the fibers of $\pi_M\colon EL_M\to L_M$.
Notice that by cutting $M$ along some $\Sigma$ as above we get $EL_M=EL_{M_2} \times_{C_\Sigma} EL_{M_1}$, so
we may also interpret $EL_M$ as a canonical correspondence, the  \textsf{evolution correspondence}. 
\end{Rem}

It is tempting to think in terms of gluing manifolds along boundary components instead of cutting them (even though this might require some extra pieces of data, like collars, or to work up to homotopy as in \cite{L}). From this perspective we could think of  Lagrangian field theories as inducing a functor from the cobordism category (of manifolds with appropriate structure) to the extended presymplectic category. We may think of this as the classical version of the Segal--Atiyah \cite{S,A} axioms for quantum field theory.

\begin{Rem}
Recall that a closed two\ndash form on a manifold $P$ is called presymplectic if its kernel is a subbundle of $TP$ (in finitely many dimensions this is equivalent to requiring
$\omega$ to have constant rank). There is no reason why the two\ndash forms $\omega_\Sigma$ we obtained above should be presymplectic, but this is a fundamental requirement
for making sense of the rest of this program. This requirement puts some constraints on the theories one can write down. 
\end{Rem}

\begin{Rem}
In the above description the two\ndash form is always exact. On the other hand, physical examples with non exact symplectic forms abound. One source for them is reduction, see the next subsection, others arise from dropping the restrictive condition that the action is a (well defined) function. More generally, one should think of the action $S$, or rather
of the Gibbs weight $\exp\frac\ii\hbar S$, as the section of a line bundle over $F_M$. In this more general setting, $\alpha_\Sigma/\hbar$ is no longer a one\ndash form on $C_\Sigma$, but
a connection one\ndash form on a line bundle.

A simple example where this occurs is that of a charged particle moving in a magnetic field on a manifold $N$. The action contains the term $\int_{t_0}^{t_1} A_i(x(t))\,\dot x^i(t)\;\dd t$, where $A=A_i dx^i$ is the vector potential regarded as a one\ndash form.
This term is also equal to $\int_\gamma A$, where $\gamma$ is the image of the path. If we make a gauge transformation,
the action then changes by boundary terms. Such action is not a function on the space of paths when $A$ is a connection on some nontrivial line bundle $E$ over $N$. Using the evaluation map at the endpoints, we can pullback this line bundle to $F_{[t_0,t_1]}$. Namely, we define
$E_{[t_0,t_1]}=\ev_{t_0}^*E^*\otimes\ev_{t_1}^*E$. We can then see the Gibbs weight as a section of $(E_{[t_0,t_1]})^{\otimes k}$ (where $k=1/\hbar$ is an integer).
The boundary one\ndash form $\alpha$ has the term $A$ as a contribution from the magnetic term in the action and therefore $\alpha/\hbar$ is defined only as a connection on $p^*E^{\otimes k}$, where
$p$ is the projection from $C=TN$ to $N$. The symplectic form $\omega$ is the canonical one for a particle on $N$ plus the curvature of $A$.


Another example is the WZW model, as discussed in \cite{Fr} \cite{G}.
In this paper, for simplicity, we will assume that the action is defined as a function.
\end{Rem}

\subsection{Reduction}
If the two\ndash form $\omega_\Sigma$ is degenerate, one may perform reduction by its kernel. If the leaf space $\underline C_\Sigma$ is smooth it inherits a symplectic structure $\underline{\omega_\Sigma}$. We may also project an evolution relation $L_M$ to the reduction and denote this by $\underline{L_M}$. If it is a smooth submanifold of $\underline{C_\Sigma}$, it is automatically isotropic. Actually, in all the examples at hand it is Lagrangian. (In example~\ref{e:nonlag}, this is trivial since the reduction of $C$ is zero\ndash dimensional.)
Finally, notice that at the reduced level $\lim_{t_1\to t_0} \underline{L_{\Sigma\times[t_0,t_1]}}$ is the graph of the identity on $\underline{C_\Sigma}$. In TFTs, this is so even without taking the limit.

\begin{Rem}
In general we cannot expect $\alpha_\Sigma$ to be horizontal with respect to the kernel of $\omega_\Sigma$, even though this happens in most examples discussed in this paper (with the notable exception of subsection~\ref{s:geod}). If this is not the case, we should regard $\alpha_\Sigma$ as a connection one\ndash form on the trivial line bundle
$E_\Sigma:=C_\Sigma\times \bbC$. Since, by definition, the restriction of this line bundle to each leaf is flat, we may reduce to a line bundle with connection
$(\underline{E_\Sigma},\underline{\alpha_\Sigma})$ if the holonomy of $\alpha_\Sigma$ is trivial on each leaf. Equivalently, we may think of
$\alpha_\Sigma$ as a contact form on the total space of $E_\Sigma$. Its reduction, if smooth, will be a contact manifold $\underline{E_\Sigma}$. Under the same conditions as above it will be the total space of a line bundle with connection over $\underline{C_\Sigma}$.
\end{Rem}

\begin{Rem}\label{r:partialred}
In general the reduced space $\underline{C_\Sigma}$ is singular and we want to avoid reduction. We will see in the next Sections how to give a good cohomological replacement for it.
However, some partial reduction is very often possible and useful. See for example Remark~\ref{r:CauchyE}. We will see several other examples in the following.
\end{Rem}


\subsection{Axiomatization}
By the above discussion we see that a Lagrangian field theory in $d$  dimensions induces the following "categorical" description:
\begin{itemize}
\item The source category is a category of cobordisms: objects are $(d-1)$\ndash manifolds and morphisms are $d$\ndash manifolds with boundary. Depending on the theory there might be restriction or additional data (e.g., a metric). Composition of morphisms is given by gluing along boundary components; one way to make sense of this consists in putting a choice of collar of the boundary in the additional data.
\item The target "category" has (usually infinite\ndash dimensional) presymplectic manifolds as its objects and correspondences with Lagrangian (or just isotropic) image as morphisms.
\end{itemize}
A few comments are in order.
\begin{enumerate}
\item What is actually important is not really gluing manifolds along common boundary, but cutting manifolds along submanifolds. This structure is more relevant than the categorical structure and much less problematic.
\item In the case of a regular field theory,
the dynamics of the problem on the space time $M$ may be recovered by choosing boundary conditions---viz., the choice of a  submanifold $L$ of the symplectic manifold $C_{\de M}$---and
take the fiber of the evolution correspondence $EL_M$ over the intersection points between $L$ and the evolution relation $L_M$ as the space of solutions for these boundary conditions.
We might require for a field theory to be good that these fibers should be generically finite dimensional; we call elements of these fibers the (classical) vacua of the theory.

In order for the variational problem to be well-defined, we have to avoid boundary terms and, as a consequence, require $L$ to be such that the restriction of
$\alpha_{\de M}$ to it vanishes. This in particular implies that $L$ must be isotropic. In order to have, generically, solutions, we should also require $L$ to be maximally isotropic, i.e., Lagrangian.
\item
In a non regular theory, boundary conditions are also given by the choice of a Lagrangian submanifold  $L$ on which the one\ndash form $\alpha$ vanishes. In the reduced theory, one considers
the intersection between the reduction of $L$ and that of the evolution relation $L_M$ and the fibers over them.
In addition, one also has to consider a reduction of these fibers. This is an additional piece of data (not contained in the Lagrangian function defining the field theory). A more refined definition of the target  category would then require endowing the evolution correspondences $EL_M$ with an integrable distribution---the \textsf{gauge symmetries} of the theory---with the consistency requirement that its image in the evolution relation
$L_M$  should coincide with  the restriction of the characteristic distribution  of the presymplectic manifold.
\item The above setting might be too rigid as the one\ndash form $\alpha_{\de M}$ might not restrict to zero on the Lagrangian submanifold $L$ one wishes to consider. To add more flexibility, one can allow changing $\alpha_{\de M}$ by an exact term $\dd f$. By consistency with \eqref{e:fundamentalFock} we see that in the Lagrangian field theory we started with
we have to change the action by $\pi_M^*f$. To preserve locality we might want $f$ itself to be a local functional.
By using Stokes Theorem, we may also write this as a bulk term. The original Lagrangian is changed by a total derivative (which is hence invisible on a manifold
without boundary).
\item In the target ``category," we might also work in the more refined setting where objects are endowed with a line bundle with connection whose curvature is the presymplectic form (a prequantization bundle).  The shift of $\alpha_{\de M}$ by an exact form in the previous comment should now be replaced by a gauge transformation for the
connection one\ndash form.
In addition, we may take care of the Hamilton--Jacobi action as a covariantly closed section of the pullback of the flat line bundle from the evolution relation to the evolution correspondence. If the fibers of the correspondence over the relation are connected, this defines a section of the flat line bundle over the relation. In many relevant examples in addition the line bundle over the presymplectic manifold is trivial; in these cases, the presymplectic form is exact and the Hamilton--Jacobi action is a function.
\item It might also make sense to allow for singular presymplectic manifolds or for singular relations/correspondences.
\end{enumerate}

\subsection{Perturbative quantization}\label{s:pertquant}
The perturbative functional integral may be extended in the presence of boundary.\footnote{What we call here perturbative perhaps should be called semiclassical. Strictly speaking the perturbative expansion would be taking
a formal power series expansion in coupling constants of the action.} Assume first that the theory is regular.
For simplicity, assume that the symplectic manifold $C_{\de M}$ is endowed with a Lagrangian foliation along which $\alpha_{\de M}$ vanishes and with a smooth leaf space $B_{\de M}$.\footnote{A more general setting would require a discussion of geometric quantization.}
Denote by $p_{\de M}$ the projection $C_{\de M}\to B_{\de M}$.
We then define the boundary vector space $H_{\de M}$ as the space of functions on $B_{\de M}$ and
the state $\psi_M$ associated to the bulk $M$ as
\[
\psi_M(\phi) = \int_{\Phi\in\pi_M^{-1}(p_{\de M}^{-1}(\phi))} \EE^{\frac\ii\hbar S_M(\Phi)}\;[D\Phi].
\]
The integral is defined by the formal saddle point approximation around critical points. As explained in the previous subsection, we may allow for some
finite\ndash dimensional degeneracy. In this case, we should think of $\psi_M$ as a function on the total space of the bundle of vacua over 
$B_{\de M}$. If this makes sense, one could also eventually perform the remaining finite\ndash dimensional fiber integration.

In non-regular theories we have two (related) problems. The first is that on the boundary space of fields
we only have a presymplectic structure. The second is that the critical points are degenerate
(with infinite\ndash dimensional fibers).
To approach this problem we have to require as an additional piece of data the choice of gauge symmetries. The idea is that the situation reduces to the regular one if we mod out by gauge symmetries in the bulk and by the characteristic foliation on the boundary. However, this usually leads to singular spaces and, even when it is not the case, one should make sense of the functional integral on the quotient. The way out is to replace reduction by its cohomological version. When $\partial M=\emptyset$ this goes under the name of BV formalisms \cite{BV} and it is known as BFV formalism \cite{BFV} in the case of boundary reduction by the characteristic foliation. The goal of this note (and of \cite{CMR}) is to show how the two formalisms fit together in a consistent way.

\subsection{An alternative approach}\label{s:alt}
Instead of introducing the space $C_\Sigma$ of Cauchy data directly, one can ``derive" it from the following construction which is somehow more natural and better fitted to the BFV formalism which will be discussed in Section~\ref{S:BFV}. The space of Cauchy data obtained in this way may not coincide with the one introduced previously, but the two construction agree after reduction.

The main idea is to associate to a $(d-1)$\ndash manifold $\Sigma$ the space $\Tilde F_\Sigma$ of germs of fields  at $\Sigma\times\{0\}$ on $\Sigma\times[0,\epsilon]$.
We will call it the space of preboundary fields. The boundary term in the variational calculus yields, as above, a one\ndash form $\Tilde\alpha_\Sigma$ on 
$\Tilde F_\Sigma$ and
the fundamental equation \eqref{e:fundamentalFock} now reads
\begin{equation}\label{e:fundamentalFocktilde}
\delta S_M = \EL_M + \Tilde\pi_M^*\Tilde\alpha_{\de M},
\end{equation}
where $\Tilde\pi_M$ is the natural surjective submersion from $F_M$ to $\Tilde F_{\de M}$.

We then introduce $\Tilde\omega_\Sigma:=\dd\Tilde\alpha_\Sigma$. This two\ndash form will have a huge kernel but is assumed to be presymplectic. We denote by
$(F^\de_\Sigma,\omega^\de_\Sigma)$ the reduced space, which we will call the space of boundary fields. For the rest of the discussion, we are going to assume $F^\de_\Sigma$ is a smooth manifold.
In general, we do not require $\Tilde\alpha_\Sigma$ to be basic. If the trivial line bundle with connection $\Tilde\alpha_\Sigma$ on $\Tilde F_\Sigma$ may be reduced to a smooth
line bundle on $F^\de_\Sigma$, we will denote by $\alpha^\de_\Sigma$ the induced connection one\ndash form.

For simplicity, we are now going to assume that
$\Tilde\alpha_\Sigma$ is indeed horizontal, so $\alpha^\de_\Sigma$ is a one\ndash form on $F^\de_\Sigma$, and leave the general case to the reader. If we denote by $\pi_M$
the composition of $\Tilde\pi_M$ with the natural projection from $\Tilde F_{\de M}$ to $F^\de_{\de M}$, we get
\[
\delta S_M = \EL_M + \pi_M^*\alpha^\de_{\de M}.
\]
Out of it we get that $L_M:=\pi_M(EL_M)$ is isotropic. We finally define the (new version of the) space of Cauchy data $C_\Sigma$ as the space of points of $F^\de_\Sigma$
that can be completed to a pair belonging to $L_{\Sigma\times[0,\epsilon]}$ for some $\epsilon$. Notice that we think of $F^\de_\Sigma$ as a relation from the one\ndash point
manifold to $F^\de_\Sigma$ and of $L_{\Sigma\times[0,\epsilon]}$ as a relation from $F^\de_\Sigma$ to itself, we may write
\[
C_\Sigma = \bigcup_{\epsilon\in(0,+\infty)} L_{\Sigma\times[0,\epsilon]}\circ F^\de_\Sigma.
\]
Notice that $F^\de_\Sigma$ is coisotropic in itself. If $\forall\epsilon$ we assume $L_{\Sigma\times[0,\epsilon]}$ also to be so---and hence to be Lagrangian---, then each composition is coisotropic (up to some infinite-dimensional subtleties), and so will be the union. However, it may happen that $C_\Sigma$ is coisotropic even if the $L$s are not Lagrangian. 

\subsubsection{Examples}
\begin{Exa}
Consider a nondegenerate Lagrangian function on $TN$ as at the beginning of Section~\ref{S:LI}. The space $\Tilde F_{pt}$ is just the infinite jet bundle over $N$. The one\ndash form is given in \eqref{e:alphaL}. The kernel of the corresponding two\ndash form consists of all jets higher than the first, so $F^\de_{pt}=TN$. Since every point in it can be completed
to a pair in $L_{[0,\epsilon]}$, for $\epsilon$ small enough, we recover $C_{pt}=TN$.
\end{Exa}

\begin{Exa}
In the case discussed in subsection~\ref{s:geod}, $\Tilde F_{pt}$ is the open submanifold in the infinite jet bundle over $\bbR^2$ obtained by requiring the first jet to be different from zero.
The reduced space $F^\de_{pt}$ is $TS^1$ with symplectic form obtained by pullback from the cotangent bundle using the metric and $L_I$ is the graph of the identity for every interval $I$; hence, $C_{pt}=F^\de_{pt}=TS^1$. Notice that in this example the space of Cauchy data given by this construction is different from the previous one, thought their reductions are obviously the same. Moreover, in this example the one\ndash form $\Tilde\alpha_{pt}$ is not basic. The induced one\ndash form connection $\alpha^\de_{pt}$ is the one discussed
in remark~\ref{r:redconn}.
\end{Exa}

\begin{Exa}
We now work out the new description of example~\ref{e:nonlag}. Here $\Tilde F_{pt}$ is the infinite jet bundle over $\bbR^2$ and $\Tilde\alpha_{pt}=yv_x\dd x$.
The kernel of the two\ndash form is given by all jets higher than the first for $x$, by all jets higher then
the zero jet for y, and by $X:=y\frac\de{\de y}-v_x\frac\de{\de v_x}$. So in this case the form is not presymplectic.
To solve this problem we assume $(v_x,y)\not=(0,0)$. This means that the original space of fields $F_I$ has to be defined as paths in $\bbR^2$ that can hit the $x$\ndash axis
only with non zero $x$\ndash velocity. The reduction is then $F^\de_{pt}=\bbR^2$. If we denote by $(p,q)$ its coordinates we have $\alpha^\de_{pt}=p\dd q$. Moreover,
$\pi_{[t_0,t_1]}(x(\cdot),y(\cdot))=(y(t_0)\dot x(t_0),x(t_0),y(t_1)\dot x(t_1),x(t_1))$. Since $EL_I$ consists of paths that are constant in the $x$\ndash direction, we get
$L_{[t_0,t_1]}=\{(0,q,0,q),\ q\in\bbR\}$ which is clearly not Lagrangian. On the other hand, we have $C_{pt}=\{(0,q),\ q\in\bbR\}$ which is coisotropic.
\end{Exa}

\begin{Exa}[Electrodynamics]\label{exa:EM}
We now discuss the case of electrodynamics (we leave to the reader the generalization to nonabelian Yang--Mills theory). The space of fields on a manifold $M$ is the space $\calA_M$
of connection
one\ndash forms for a fixed line bundle over $M$. The action is
$S_M(A)=(\dd A,\dd A)=\int_M dA\wedge *dA$, where $(\ ,\ )$ is the Hodge pairing of forms for a fixed metric on $M$
and $*$ is the Hodge $*$-operation. The EL equations are $\dd^*\dd A=0$, where $\dd^*$ is the formal adjoint of $\dd$
with respect to the Hodge pairing.

For simplicity of exposition, we now reformulate electrodynamics in the first-order formalism (and leave to the reader its study in the usual second-order formalism). Namely,
we enlarge the space of fields to $F_M:=\calA_M\times\Omega^{d-2}(M)$ and extend the action to
\[
S_M(A,B)=\int_M B\wedge \dd A + \frac12 B\wedge \,*B
\]
The EL equations are $\dd A + *B=0$ and $\dd B=0$. Hence, $B$ is completely determined by $A$ and then $A$ must satisfy $\dd^*\dd A=0$.

On the space of preboundary fields $\Tilde F_\Sigma$ we have the one\ndash form $\Tilde\alpha_\Sigma=\int_\Sigma B\wedge \delta A$.
It then follows that the space of boundary fields is $F^\de_\Sigma=\calA_\Sigma\times\Omega^{d-2}(\Sigma)$ with one\ndash form
$\alpha^\de_\Sigma=\int_\Sigma B\delta A$ and symplectic form $\omega^\de_\Sigma=\dd\alpha^\de_\Sigma$.

Let us now consider $L_{\Sigma\times[0,\epsilon]}$. The equation $\dd B=0$ restricts to the boundary, so it has to be satisfied by a field in $L_{\Sigma\times[0,\epsilon]}$. We will call the direction along $[0,\epsilon]$ vertical.
The evolution will impose some other conditions, but we claim that $C_{\Sigma}=\calA_\Sigma\times\Omega^{d-2}_{closed}(\Sigma)$, namely that no other conditions have to be imposed on the first boundary component. The reason is that $A$ on the first boundary can always be extended to a solution. In the axial gauge (i.e., when we require that the vertical component of $A$ vanishes), the solution is unique once we specify the vertical derivative of $A$ at the first boundary. But this first derivative may be chosen, actually uniquely, so as to yield the given closed $B$ on the boundary.

Notice that $B$---despite the notation---is the electric field on the boundary (or, better, the $(d-1)$\ndash form corresponding to the electric vector field using the metric) and the equation $\dd B=0$ is just the Gauss law (i.e., divergence of electric field equal to zero). The characteristic distribution on $C_\Sigma$ consists of just the gauge transformations
on $A$.
\end{Exa}

\begin{Exa}[Abelian $BF$ theories]\label{exa:abeBF}
We may consider the ``topolological" limit of the first-order formulation of electrodynamics, i.e., drop the term with the Hodge $*$ operator. This way we get the action
$S_M=\int_M B\wedge \dd A$ on the same space of fields $F_M=\calA_M\times\Omega^{d-2}(M)$. This theory is called abelian $BF$ theory.

The space of boundary fields is the same as for electrodynamics. What changes are the Lagrangian submanifolds $L_M$. Since the EL equations are just $\dd A=0$
and $\dd B=0$, we see that $L_M$ consists closed $A$ and $B$ on the boundary that can be extended to closed $A$ and $B$ in the bulk. As a result
$C_\Sigma=\calA_\Sigma^{flat}\times \Omega^{d-2}_{closed}(\Sigma)$. The characteristic distribution consists of gauge transformations for $A$ and shifts of $B$ by exact forms.
\end{Exa}

\section{The BFV formalism}\label{S:BFV}
In this Section we address the problem of reformulating the reduction of a presymplectic manifold $C$ cohomologically.

If we work in the setting of subsection~\ref{s:alt}, our presymplectic submanifold is actually given as a coisotropic submanifold of a symplectic manifold. Otherwise,
we first recall that Gotay \cite{Go} proved that
every presymplectic manifold $(C,\omega_C)$ may be embedded into a symplectic manifold $(F,\omega_F)$ as a coisotropic submanifold such that $\omega_C$ is the restriction of
$\omega_F$ to $C$. Moreover, such an embedding is unique up to neighborhood equivalence. The existence part is simply proven by taking $F=D^*$, where $D$ is the kernel
of $\omega_C$, and $\omega_F=p^*\omega_C+\sigma^*\omega_\text{can}$, where $p$ is the projection $D^*\to C$, $\omega_\text{can}$ is the canonical symplectic form on
$T^*C$ and $\sigma$ is a splitting of $T^*C\to D^*$. Notice that $\omega_\text{can}$ is exact; so an exact presymplectic manifold can be coisotropically embedded into an exact symplectic manifold\footnote{ In this note
we will not focus on subtleties of this statement in the infinite dimensional setting.}.

We are then led to consider the problem of how to describe symplectic reduction of a coisotropic submanifold cohomologically. This goes under the name of BFV
formalism \cite{BFV}.

We start with the finite dimensional setting.
Locally, a coisotropic submanifold $C$ of $F$ can be described as the common zero locus of some differentiably independent functions $\phi_i$ on $F$.
The characteristic foliation is then the span of the Hamiltonian vector fields $X_i$ of the $\phi_i$s. The space of functions on the quotient $\underline C$, if it is smooth,
is the same as $(C^\infty(F)/\langle \phi_1,\phi_2,\dots \rangle)^{(X_1,X_2,\dots)}$, where $\langle \phi_1,\phi_2,\dots \rangle$ denotes the ideal generated by the $\phi_i$s and
the exponent denotes taking the subalgebra invariant under all the $X_i$s. The goal is to describe this space (actually this Poisson algebra) as the zeroth cohomology of a complex
(actually the differential graded Poisson algebra of functions on a graded symplectic manifold).

To do this we add to $F$ new odd coordinates $b_i$ of degree $-1$ and define a vector field $Q$ on the supermanifold so obtained by imposing $Q(f)=0$ for any $f\in C^\infty(F)$
and $Q(b_i)=\phi_i$. The cohomology is concentrated in degree zero and yields $C^\infty(F)/\langle \phi_1,\phi_2,\dots \rangle$. To select the invariant part,
we add more odd coordinates $c^i$ of degree $+1$ and extend $Q$ to the  supermanifold so obtained $\calF$
by requiring $Q(b_i)=\phi_i$, $Q(c^i)=0$ and $Q(f)=c^iX_i(f)$. This $Q$ is a Hamiltonian vector field on $\calF$ with respect to the symplectic form
$\omega_F+\dd b_i\,\dd c^i$ and Hamiltonian function $c^i\phi_i$; yet, in general it is not  a differential on $C^\infty(\calF)$. However, using cohomological perturbation theory one can prove \cite{StaBFV} that 
the Hamiltonian function may be deformed in such a way this occurs. The construction may also be globalized \cite{Bor,Her,SchBFV}. 
To summarize, we have the
\begin{Thm}
Let $C$ be a coisotropic submanifold of a finite\ndash dimensional symplectic manifold $F$. Then one can embed $F$ as the body of a supermanifold $\calF$
with an additional $\bbZ$\ndash grading endowed with an even symplectic form $\omega_\calF$ of degree zero and an even function $S$ of degree $+1$
such that its Hamiltonian vector field $Q$  squares to zero and its
cohomology in degree zero is isomorphic as a Poisson algebra  to the algebra of functions on $C$ that are invariant under its characteristic distribution.
This construction is unique up to symplectomorphisms of $\calF$ if one requires it to be minimal (in terms of the newly added coordinates).
\end{Thm}

In the case of field theory, the analogous result---with the additional condition that $S$ and $\omega_\calF$ are local---was
proved long ago by \cite{BFV} (in the description above the index $i$ is now replaced by a worldsheet coordinate and the sum over $i$ by an integral).
Notice however that, in order to get $S$ as a local functional, one often has
to add extra fields of degree greater than $+1$ (and consequently extra fields of degree less than $-1$). In any case, the final result is what we will call a BFV manifold.
\begin{Def}
A BFV manifold is a triple $(F,\omega,Q)$ where $F$ is a supermanifold with additional $\bbZ$\ndash grading, $\omega$ is an even symplectic form of degree zero,
and $Q$ is an odd symplectic vector field of degree $+1$ satisfying $[Q,Q]=0$.
\end{Def}
\begin{Rem}\label{r:QBFV}
Recall that $Q$ symplectic means $L_Q\omega=0$. On the other hand the $\bbZ$\ndash grading amounts to the existence of an even vector field $E$ of degree zero (the graded Euler vector field)  such that the grading on functions, forms and vector fields is given by the eigenvalues of the Lie derivative $L_E$. We then have
$L_E\omega=0$ and $[E,Q]=Q$. This then implies that $Q$ is automatically Hamiltonian, $\iota_Q\omega=\dd S$, with $S=\iota_E\iota_Q\omega$ (this remark is due to Roytenberg \cite{R}). Notice that the condition $[Q,Q]=0$ implies the ``classical master equation" (CME) in the BFV formalism:
\[
\{S,S\}=0,
\]
where $\{\,\ , \, \}$ denotes the Poisson bracket induced by $\omega_F$.
\end{Rem}

The coisotropic submanifold $C$ can also be recovered geometrically. Namely, one defines $\calEL$ as the zero locus of $Q$. More precisely (when $EL$ is singular as is often the case), one considers the ideal $I_{\calEL}$ generated by functions of the form $\{S,f\}$ with $f\in C^\infty(\calF)$. This ideal is clearly a Lie subalgebra (with respect to the Poisson bracket) thanks to the CME. This amounts to saying that $\calEL$ is a coisotropic submanifold. The original $C$ is just its body.

\begin{Rem}[Quantization]
The (geometric) quantization of $F$ is in this setting replaced by a quantization of $\calF$ with a compatible quantization of $S$. Namely, one has to produce a graded vector space $\calH_\calF$ quantizing $(\calF,\omega_\calF)$ together with an odd operator $\Omega$ of degree $1$ quantizing $S$ and satisfying $\Omega^2=0$.
Notice that the CME is the classical counterpart of the last equality and there might be obstruction (``anomalies") in finding such an $\Omega$. If everything works, however, one can
consider the cohomology of $\Omega$. Its degree zero component may be thought of the quantization of the reduction of $C$.
\end{Rem}

\subsection{BFV as a boundary theory}
In Section \ref{S:LII} we saw that a $d$\ndash dimensional Lagrangian field theory associates a space $C_\Sigma$ with a closed (often exact) two\ndash form $\omega_\Sigma$ to a $(d-1)$\ndash dimensional manifold $\Sigma$. It was part of the assumptions that $C_\Sigma$ is a manifold and that $\omega_\Sigma$ is presymplectic. Following the description above, we now associate to $\Sigma$ a BFV manifold $(\calF^\de_\Sigma,\omega^\de_\Sigma,Q^\de_\Sigma)$. (The upper symbol $\de$ is a reminder that this is the boundary construction as in the following we will have a similar construction, with similar notations, for the bulk.) If we work in the settings of subsection~\ref{s:alt},
then we take $F^\de_\Sigma$ to be the degree zero part of $\calF^\de_\Sigma$.




An important remark is that other cohomology groups may turn out to be nontrivial. As an example, consider first-order electrodynamics as in example~\ref{exa:EM}.
Recall that the space of boundary fields on $\Sigma$ is $\calA_\Sigma\times\Omega^{d-2}(\Sigma)$ whereas $C_\Sigma=\calA_\Sigma\times\Omega^{d-2}_{cl}(\Sigma)$.
To implement the BFV construction we add odd fields $c\in\Omega^0(\Sigma)$ of degree $+1$ and $b\in\Omega^{d-1}(\Sigma)$ of degree $-1$ and consider the BFV action
$\calS_\Sigma=\int_\Sigma c\dd B$. The Hamiltonian vector field $Q$ acts trivially on $c$ and $B$. On the other hand $Qb=\dd B$ and $QA=\dd c$.
So the BFV cohomology yields functions on $H^0(\Sigma)[1]\times \calA_\Sigma^{flat}/{gauge} \times \Omega^{d-2}_{cl}(\Sigma)\times H^{d-1}(\Sigma)[-1]$.
The extra factors, in degree $0$ and $d-1$, express the stacky nature of the reduction and become even more important in the nonabelian Yang--Mills case.

\section{The BV formalism for manifolds with boundary}\label{S:BV}
The BV formalism \cite{BV} deals with the degeneracy problem for an action in the bulk. In the BV case, as in Sections~\ref{S:LI} and \ref{S:LII}, we have a $d$\ndash dimensional
Lagrangian field theory, i.e. the assignment of a space of fields $F_M$ and an action $S_M=\int_M L$ over $F_M$ to a $d$\ndash manifold $M$. But in addition we have
a distribution $D_M\subset T F_M$ on $F_M$ which describe the ``symmetries'.
This distribution does not have to be given by an action of a Lie
group. It is involutive and of finite codimension when restricted to $EL_M$. The construction aims at cohomologically resolving the quotient of $EL_M$ by the symmetries. Let us sketch the last point assuming at the beginning that our manifold $M$ has no boundary. The BV construction proceeds by
\begin{enumerate}
\item first extending the space of fields $F_M$ to the supermanifold $D_M[1]\subset T[1] F_M$ (i.e., one assigns degree $1$ and the according Grassmann parity to coordinates on fibers of $D_M$),
\item extending the action $S_M$ to a new local functional $\calS_M$ on $\calF_M:=T^*[-1]D_M[1]$ which has the two following properties:
\begin{enumerate}
\item It satisfies the classical master equation (CME) $\{\calS_M,\calS_M\}=0$, where $\{\ ,\ \}$ is the (degree $+1$) Poisson bracket associated to the canonical symplectic form (of degree $-1$)
on $\calF_M$, and
\item the restriction of $D_M$ to $EL_M$ is the same as the restriction of the characteristic distribution of the coisotropic submanifold $\calEL_M$ of critical points of $\calS_M$
to its degree zero part.
\end{enumerate}
\end{enumerate}
The solution can be found by cohomological perturbation theory. In order to preserve locality, it is often necessary to extend the above procedure by allowing
dependent symmetries and resolving their relations by adding new variables of degree $2$ (ghosts for ghosts) and so on. The final result is anyway a
supermanifold with odd symplectic form of degree $-1$ and a solution of the CME.

\begin{Rem}
The CME is also the starting point for making sense of the integral of $\EE^{\frac\ii\hbar S_M}$ over $F_M$. In the saddle point approximation, one expands around critical points, i.e., points of $EL_M$.
If the action is degenerate---namely, its Hessian at a critical point  is degenerate---one cannot even begin the perturbative expansion. However, if one quotients by a distribution as above, one saves the game (or at least reduces the problem to a residual finite dimensional integration). This quotient might be very singular; also notice that in general situations the distribution is not even involutive outside of $EL_M$; and even if everything worked out properly, it might be difficult to define the perturbative functional integral on the quotient, which might have a much more involved manifold structure. The way out is to extend $S_M$ to a (possibly $\hbar$\ndash dependent) solution $\Tilde\calS_M$ of the quantum master equation (QME) on $\calF_M$. Namely, one picks a Berezinian $\rho$ (formally, since we are working in an infinite dimensional context) on $\calF_M$ and defines the
BV Laplacian $\Delta$ by $\Delta f=\frac{1}{2}\Div_\rho X_f$, where $X_f$ is the Hamiltonian vector field of a function $f$ and $\Div_\rho$ is the divergence operator with respect to $\rho$.
One requires $\rho$ to restrict to the original measure on $F_M$ and to be compatible with the symplectic structure: namely, one requires $\Delta^2=0$. The QME then reads
$\frac12 \{\Tilde \calS_M,\Tilde \calS_M\}-\ii\hbar\Delta \Tilde \calS_M=0$. The limit of $\Tilde \calS_M$  for $\hbar\to0$ solves the CME and is taken to be $\calS_M$.
One actually starts with $\calS_M$ and tries to extend it to a formal power series in $\hbar$ that solves the QME if there are no obstructions (``anomalies").
A consequence of the QME is that the integral of $\EE^{\frac\ii\hbar \Tilde \calS_M}$ on a Lagrangian submanifold is invariant under deformations of the Lagrangian submanifold.
One then replaces the originally ill-defined integral over $F_M$ by the integral over a deformation of the Lagrangian submanifold $D_M[1]$ where it is well-defined. We refer to \cite{Schw} for a good introduction.
\end{Rem}

\begin{Def} A BV manifold is a triple given by a supermanifold with additional $\bbZ$\ndash grading, an odd symplectic form of degree $-1$ and a function of degree $0$ that satisfies the CME, i.e. Poisson commutes with itself.
\end{Def}

We may then formulate the result of the BV construction in $d$\ndash dimensional Lagrangian field theory
as the  assignment of a BV manifold $(\calF_M,\omega_M,\calS_M)$ to a $d$\ndash manifold $M$.
Notice that as a consequence of the CME the Hamiltonian vector field $Q_M$ of $\calS_M$,
\begin{equation}\label{e:QoS}
\iota_{Q_M}\omega_M = \delta \calS_M,
\end{equation}
 is cohomological, i.e., it satisfies $[Q_M,Q_M]=0$.

\subsection{The case with boundary}\label{s:casebry}
Now let us allow $M$ to have a nonempty boundary. Since the BV construction is local it still assigns to $M$ a quadruple $(\calF_M,\omega_M,\calS_M,Q_M)$.
It is still true that $\omega_M$ is symplectic and that $Q_M$ is cohomological. On the other hand, $S_M$ is no longer its Hamiltonian. The problem is that \eqref{e:QoS}
involves integration by parts. We may overcome this problem working as in the previous Sections (in particular, subsection~\ref{s:alt}).

Namely, we define the space $\Tilde\calF_\Sigma$ of preboundary fields on a $(d-1)$\ndash manifold $\Sigma$ as the germs at $\Sigma\times\{0\}$ of
$\calF_{\Sigma\times[0,\epsilon]}$. Integration by parts in the computation of $\delta\calS_{\Sigma\times[0,\epsilon]}$ yields a one\ndash form
$\Tilde\alpha_\Sigma$ on $\Tilde\calF_\Sigma$. We denote by $\Tilde\omega_\Sigma$ its differential---which we assume to be presymplectic---and by
$(\calF^\de_\Sigma,\omega^\de_\Sigma)$ its reduction. We also assume that $\Tilde\alpha_\Sigma$ reduces to a connection one\ndash form $\alpha^\de_\Sigma$
on $\calF^\de_\Sigma$. In most examples, $\alpha^\de_\Sigma$ will be an actual one\ndash form.

If we take care of boundary terms, instead of \eqref{e:QoS}
we now get
\begin{equation}\label{e:QoSTilde}
\iota_{Q_M}\omega_M = \delta \calS_M +\Tilde\pi_M^*\Tilde\alpha_{\de M},
\end{equation}
where $\Tilde\pi_M$ is the natural surjective submersion from $\calF_M$ to $\Tilde\calF_{\de M}$. If we denote by $\pi_M$ the composition of $\Tilde\pi_M$
with the natural surjective submersion from $\Tilde\calF_{\de M}$ to $\calF^\de_{\de M}$, we finally get the fundamental equation of the BV theory for manifolds with boundary
\cite{CMR}:
\begin{equation}\label{e:QoSde}
\iota_{Q_M}\omega_M = \delta \calS_M +\pi_M^*\alpha^\de_{\de M},
\end{equation}

To complete the description of the theory with boundary, we still have to study $Q_M$.
The first obvious remark is that it is $\Tilde\pi_M$\ndash projectable. More precisely, for every $\Sigma$, there is a uniquely defined vector field $\Tilde{Q}_\Sigma$
(automatically cohomological)  on
$\Tilde\calF_\Sigma$ such that for every $M$ the vector field $Q_M$ projects to $\Tilde Q_{\de M}$: namely,
$\Tilde Q_{\de M}(\phi)=\dd_{\Hat\phi}\Tilde\pi_M(Q_M(\Hat\phi))$, $\forall\phi\in\Tilde\calF_{\de M}$ and $\forall\Hat\phi\in\Tilde\pi_M^{-1}(\phi)$.

Let us now differentiate \eqref{e:QoSTilde}. Using the fact that $\omega_M$ is closed, we get $L_{Q_M}\omega_M=\Tilde\pi_M^*\Tilde\omega_{\de M}$ (which by the
way proves that $Q_M$ is not even symplectic). We now apply $L_{Q_M}$ to this equation. Using the fact $Q_M$ is cohomological and projectable, we get
$\Tilde\pi_M^*L_{\Tilde Q_{\de M}}\Tilde\omega_{\de M}=0$. Since $\Tilde\pi_M$ is a surjective submersion, we conclude that $L_{\Tilde Q_{\de M}}\Tilde\omega_{\de M}=0$.

Actually, this proves that, for every $\Sigma$, $\Tilde\omega_\Sigma$ is $\Tilde Q_\Sigma$\ndash invariant. This implies that $\Tilde Q_\Sigma$ is projectable to the reduction.
To show this, we have just to check that $[\Tilde Q_\Sigma,X]$ belongs to the kernel of $\Tilde\omega_\Sigma$ for every $X$ in the kernel. This follows
from the identities $\iota_{[\Tilde Q_\Sigma,X]}\Tilde\omega_\Sigma = [L_{\Tilde Q_\Sigma}, \iota_X] \Tilde\omega_\Sigma = 0$.

We conclude that, for every $\Sigma$, there is a uniquely defined vector field $Q^\de_\Sigma$ on $\calF^\de_\Sigma$ (automatically cohomological and symplectic) to which
$\Tilde Q_\Sigma$ projects. This has two fundamental consequences:
\begin{enumerate}
\item To each $(d-1)$\ndash dimensional manifold $\Sigma$, we now associate a BFV manifold $(\calF^\de_\Sigma, \omega^\de_\Sigma,Q^\de_\Sigma)$, see definition 4.2.
\item For each $d$\ndash manifold $M$, $Q_M$ $\pi_M$\ndash projects to $Q^\de_M$.
\end{enumerate}
These two final observations together with \eqref{e:QoSde} constitute the framework of the BV formalism extended to manifolds with boundaries \cite{CMR}, which we call the BV-BFV formalism.

\begin{Def}
We define a BV-BFV manifold over a given exact BFV manifold $(\calF^\de,\omega^\de=\dd\alpha^\de,Q^\de)$ as a quintuple
$(\calF,\omega,\calS,Q,\pi)$ where $\calF$ is a supermanifold with additional $\bbZ$\ndash grading, $\omega$ is an odd symplectic form of degree $-1$,
$\calS$ is an even function of degree $0$, $Q$ is a cohomological vector field and $\pi\colon\calF\to\calF^\de$ is a surjective submersion such that
\begin{enumerate}
\item $\iota_Q\omega = \dd\calS + \pi^*\alpha^\de$,
\item $Q^\de=\dd\pi Q$.
\end{enumerate}
The definition may be extended to BV-BFV manifolds over a BFV manifold with connection $\alpha^\de$. This requires introducing a line bundle over
$\calF^\de$ and viewing $\exp(\frac{i}{\hbar}\calS)$ as a section of the pulled-back bundle.
\end{Def}

\begin{Rem}[Axiomatization]
We may now reformulate Lagrangian field theories axiomatically as a "functor" from some "cobordisms category" to a category where
the objects are BFV manifold with connection and the morphisms are BV-BFV manifolds over the Cartesian product of the objects.
\end{Rem}

\begin{Rem}
Notice that using this method the BFV construction associated to the boundaries is obtained from the BV construction in the bulk and does not have to be done independently.
Also recall that, by general principles \cite{R}, $Q^\de_\Sigma$ is Hamiltonian with a uniquely defined odd Hamiltonian function $\calS^\de_\Sigma$ of degree $+1$. This yields as a consequence the following generalization of the CME
\begin{equation}\label{e:modCME}
Q_M(\calS_M) = \pi^*_M(2\calS^\de_{\de M} -\iota_{Q^\de_{\de M}}\alpha^\de_{\de M}),
\end{equation}
which can be proved as follows. First differentiate \eqref{e:QoSde} to obtain $L_{Q_M}\omega_M=\pi_M^*\omega^\de_{\de M}$.
Then apply $\iota_{Q_M}$ to \eqref{e:QoSde} and use the obtained equation and the fact that $Q_M$ is cohomological and projects to
$Q^\de_{\de M}$ to obtain the differential of \eqref{e:modCME}. Then observe that the differential of a function of degree $1$ vanishes if and only if the function itself vanishes (we have no constants in degree $+1$).
\end{Rem}

\begin{Exa}[First-order electrodynamics]\label{exa:EMBV}
We return to example~\ref{exa:EM} (first-order YM is explained in details in \cite{CMR};
we leave the usual second-order formulation as an exercise to the reader).
Since we want to implement gauge transformations for $A$, we define $D_M[1]$ by adding the ``ghost field" $c\in\Omega^0(M)$ which is odd and of
degree $+1$. Gauge transformations are given by the vector field $\dd c$ in the $A$\ndash direction (here $\dd$ denotes the de~Rham differential
on $M$).
The BV space of fields
$\calF_M=T^*[-1]D_M[1]$ is then
\[
\Omega^0(M)[1]\times\calA_M\times\Omega^2(M)[-1]\times\Omega^{d-2}(M)\times\Omega^{d-1}(M)[-1]\times\Omega^d(M)[-2].
\]
We add a superscript $+$ to denote the canonically conjugate coordinates (a.k.a.\ antifields) to the fields: namely,
$B^+\in\Omega^2(M)[-1]$, $A^+\in\Omega^{d-1}(M)[-1]$, $c^+\in\Omega^d(M)[-2]$.
The BV action is just
\[
\calS_M=\int_M B\wedge \dd A + \frac12 B\,\wedge *B +A^+\wedge \dd c.
\]
The cohomological vector field $Q_M$ acts as follows (we omit the terms where the action is zero):
\[
Q_MA=\dd c,\ Q_MA^+=\dd B,\ Q_M B^+=*B+\dd A,\ Q_Mc^+=\dd A^+.
\]
On the space of preboundary fields on a $(d-1)$\ndash manifold $\Sigma$, we get
$\Tilde\alpha_\Sigma=\int_\Sigma B\,\wedge \delta A + A^+\,\delta c$. We immediately see that the kernel of $\Tilde\omega_\Sigma$
consists of all jets of $B^+$ and $c^+$ and all jets higher then the zeroth of $A,B,A^+,c$. Moreover, $\Tilde\alpha_\Sigma$ is also basic.
We then get
\[
\calF^\de_\Sigma=\Omega^0(\Sigma)[1]\times\calA_\Sigma\times\Omega^{d-2}(\Sigma)\times\Omega^{d-1}(\Sigma)[-1].
\]
Projecting the cohomological vector field, we get the cohomological vector field $Q^\de_\Sigma$, which acts by
$Q^\de A^+=\dd B$ and $Q^\de A = \dd c$ and has Hamiltonian function $S^\de_\Sigma = \int_\Sigma c\,\dd B$.
\end{Exa}



\subsection{EL correspondences}
We now define the space $\calEL_M$ as the space of zeros of $Q_M$ and $\calL_{\de M}$ as its image under $\pi_M$. This generalizes the classical
story of evolution correspondences and evolution relations. Notice that \eqref{e:QoSde} implies that $L_{\de M}$ is isotropic, and we are going to assume that it is actually Lagrangian. There are two problems to be tackled though: the first is that $\calEL_M$ is not smooth in general, the second is that we are usually interested in reduction, which is even more singular in general.

One way to avoid the first problem is by working with an algebraic description only, but we will often pretend that we are dealing with smooth manifolds.
Namely, instead of $\calEL_M$ we consider its vanishing ideal
$I_{\calEL_M}$, i.e., the ideal generated by functions of the form $Q_Mf$, with $f\in C^\infty(\calF_M)$. This ideal is a Lie algebra with respect to the Poisson
bracket, which amounts to saying that $\calEL_M$ is coisotropic. If $M$ has no boundary, this is obvious since $Q_M$ is symplectic and squares to zero. If $M$ has a boundary,
this is still true since, to generate it, it is enough to consider functions $f$ that ``vanish near the boundary" (namely, functions in $\pi_{M,U}^*\calF_U$
where $U$ is compact in the interior of $M$ and $\pi_{M,U}$ is the restriction map).
The characterstic distribution $\calD_M$ is generated by the Hamiltonian vector fields of functions of the form $Qf$. If $f$ is as above, we have $Qf=\{S,f\}$ and hence
the characteristic distribution is generated by vector fields of the form $[Q,X]$ where $X$ is a Hamiltonian vector field vanishing near the boundary (we assume here that components of $Q$ on $EL$ are differentiably independent). The reduction
$\underline{\calEL_M}$ of $\calEL_M$ by $\calD_M$ carries again a symplectic form of degree $-1$ (if it is singular, this has to made sense of; e.g., by considering the open smooth locus or using the language of derived algebraic geometry \cite{PTV}).

If $M$ has a boundary, it makes sense to consider another reduction, namely by the distribution $\calD_M^Q\supset\calD_M$ generated by vector
fields of the form $[Q,X]$ where $X$ is Hamiltonian (but with no vanishing condition).
More precisely, observe that, since $Q_M$ projects to $Q^\de_{\de M}$, we have that $\calL_{\de M}$
is contained in $\calEL^\de_{\de M}$, the space of zeros of $Q^\de_{\de M}$, which is also coisotropic. Hence its characteristic distribution, generated by
vector fields of the form $[Q^\de_{\de M}, X]$ with $X$ Hamiltonian, is tangent to $\calL_{\de M}$. Now let $\ell$ be a point in $\calL_{\de M}$ and
let $[\ell]$ denote its orbit. Then $\calD_M^Q$ restricts to $\pi_M^{-1}([\ell])\cap\calEL_M$ and we denote by $\underline{\calE}_{[\ell]}$ its quotient.
The union of the $\underline{\calE}_{[\ell]}$s over $[\ell]\in\underline{\calL_{\de M}}$ is the quotient $\underline{\calEL_M}_Q$
of $\calEL_M$ by $\calD_M^Q$, which is by itself
a quotient of the symplectic reduction $\underline{\calEL_M}$.

In \cite{CMR} it is shown
that each fiber $\underline{\calE}_{[\ell]}$ carries a symplectic form of degree $-1$. This follows from a different but equivalent description of this quotient.
Namely, pick a Lagrangian submanifold $\calL$ of $\calF^\de_{\de M}$ transversal to $\calL_{\de M}$ at $\ell$. Then one shows that
$\pi_M^{-1}(\ell)\cap\calEL_M$ is coisotropic in $\pi_M^{-1}(\calL)$ and that its reduction is $\underline{\calE}_{[\ell]}$.

We call $\underline{\calE}_{[\ell]}$ the moduli space of vacua at $[\ell]$ and we assume it to be finite dimensional (if this is not the case, it means that we have not considered enough symmetries).

Notice that there is in principle a second (usually coisotropic) submanifold
$\calC'_\Sigma$ of $\calF_\Sigma$. Namely, the elements of $\calF_\Sigma$ are zeroes of $Q^\de_\Sigma$ which can be
extended as zeroes of $Q_{\Sigma \times [0,\epsilon]}$
for some $\epsilon>0$. It is meaningful to require $\calC'_\Sigma=\calC_\Sigma$. Otherwise, it again means that
we have not taken enough symmetries into account.

\begin{Rem}[Axiomatization]
If reduction were always nice, we could get the following induced axiomatization of a $d$\ndash dimensional Lagrangian field theory in the BV-BFV formalism.
To a $(d-1)$\ndash manifold $\Sigma$ we associate a symplectic supermanifold $\underline{\calEL^\de_{\partial M}}$ and to a $d$\ndash manifold $M$
we associate the ``evolution correspondence" $\underline{\calEL_M}_Q\to\underline{\calEL^\de_\Sigma}$ which has a Lagrangian image and
whose fibers are finite dimensional symplectic manifolds in degree $-1$.
If we cut a manifold $M$ along a submanifold $\Sigma$, we may try to recover $\underline{\calEL_M}$ out of the composition of the evolution correspondences, and some more data, for the two halves. This problem started to be addressed in \cite{CMR}.
\end{Rem}

\begin{Exa}[Electrodynamics---continued]
In \cite{CMR}, to which we refer for details, it is shown that, in the case of first-order electrodynamics, for any $\ell$ we have
\[
\underline{\calE}_{[\ell]}\simeq H^1(M,\de M)\oplus H^{n-1}(M)[-1]\oplus H^0(M,\de M)[1]\oplus H^n(M)[-2],
\]
which is indeed finite-dimensional. (Here $H^*(M,\de M)$ denotes cohomology relative to the boundary.)
\end{Exa}

\subsection{Extended theories}
The construction in subsection~\ref{s:casebry} may be applied iteratively to go to lower and lower dimension.
Namely, there we have obtained a BFV structure $(\calF^\de_\Sigma,\omega^\de_\Sigma,Q^\de_\Sigma)$, to which we canonically associate
a function $\calS^\de_\Sigma$, for every $(d-1)$\ndash dimensional manifold $\Sigma$ without boundary; yet, since the construction is local, we can use these data
on a $(d-1)$\ndash dimensional manifold $\Sigma$ with boundary. Again what is not going to work is the condition that $\calS^\de_\Sigma$ is the Hamiltonian
function of $Q^\de_\Sigma$. We correct this equation using the induced one\ndash form on the space of preboundary fields on $\de\Sigma$ and reduce
by the kernel of the two form. Since $\calS^\de_\Sigma$ has degree $1$, this will also be the degree of the induced symplectic form.

As a result, to a $(d-2)$\ndash manifold $\gamma$ we associate a triple
$(\calF^{\de\de}_\gamma,\omega^{\de\de}_\gamma=\dd\alpha^{\de\de}_\gamma,Q^{\de\de}_\gamma)$, where $\omega^{\de\de}_\gamma$ is an odd
symplectic form of degree $+1$ and $Q^{\de\de}_\gamma$ is a cohomological, symplectic vector field (hence automatically Hamiltonian with a uniquely
defined even Hamiltonian function $\calS^{\de\de}_\gamma$ of degree $+2$). To a $(d-1)$\ndash manifold $\Sigma$ with boundary
we now associate a quintuple $(\calF^\de_\Sigma,\omega^\de_\Sigma,Q^\de_\Sigma,\calS^\de_\Sigma,\pi^\de_\Sigma)$,
where $\omega^\de_\Sigma$ is an even symplectic form of degree $0$, $Q^\de_\Sigma$ is a cohomological vector field, $\calS^\de_\Sigma$
is an odd function of degree $+1$ and $\pi^\de_\Sigma\colon\calF^\de_\Sigma\to\calF^{\de\de}_{\de\Sigma}$ is
 a surjective submersion such that
\begin{enumerate}
\item $\iota_{Q^\de_\Sigma}\omega^\de_\Sigma = \dd\calS^\de_\Sigma + (\pi^\de_\Sigma)^*\alpha^{\de\de}_{\de\Sigma} $,
\item $Q^{\de\de}_{\de\Sigma}=\dd\pi Q^\de_\Sigma$.
\end{enumerate}
We can now consider the zero locus $\calEL^{\de\de}_{\de\Sigma}$ of $Q^{\de\de}_{\de\Sigma}$, which is coisotropic and contains
$L^\de_\Sigma:=\pi^\de_\Sigma(\EL^\de_\Sigma)$. Repeating the same analysis as above, we conclude that
$\underline{\calEL^\de_\Sigma}_Q\to\underline{\calEL^{\de\de}_{\de\Sigma}}$ has fibers with a symplectic structure in degree zero.
Notice that in general these fibers will not be finite dimensional (it would be too restrictive to ask for that).

The construction may now be iterated to $(d-2)$\ndash manifolds with boundaries. Every time the degree of the symplectic form and of the action increase by $1$. However, it probably makes sense to continue this construction only as long as the $\underline{\calE^{\de\de\dots\de}}_{[\ell]}$ spaces
are finite dimensional.

Typically, at some point we get $\calS^{\de\de\dots\de}=0$, so that $\underline{\calE^{\de\de\dots\de}}_{[\ell]}$ is the whole space of fields in the bulk
over a point on the boundary and this will usually be infinite dimensional.

On the other hand, in topological field theories of the AKSZ type \cite{AKSZ}, this construction can be iterated down to dimension $0$ always
with finite dimensional $\underline{\calE^{\de\de\dots\de}}_{[\ell]}$ fibers; see \cite{CMR} for details. Hence we can speak of fully extended field theories.
These looks like the BV-BFV version of \cite{L}.

\begin{Exa}[Electrodynamics---continued]
In example~\ref{exa:EMBV}, we got the BFV structure for first-order electrodynamics.
Applying the above reasoning, we first consider the space of preboundary fields with one\ndash form
$\Tilde\alpha^{\de\de}_\gamma=\int_\gamma B\,\delta c$. The kernel of its differential consists of all jets for $A$ and $A^+$ and
and all jets higher than the zeroth for $B$ and $c$, and $\Tilde\alpha^{\de\de}_\gamma$ is basic. Hence we get
\[
\calF^{\de\de}_\gamma=\Omega^0(\gamma)[1]\times\Omega^{d-2}(\gamma).
\]
One can also easily realize that $Q^{\de\de}_\gamma=0$. Moreover, one can also compute, see \cite{CMR},
\[
\underline{\calEL^\de_\Sigma}_Q=
\Omega^1(\Sigma)/\Omega^1(\Sigma)_{\text{\tiny exact }}\oplus\Omega_\text{closed}^{d-2}(\Sigma,\de\Sigma)\oplus H^0(\Sigma,\de \Sigma)[1]\oplus H^{d-1}(\Sigma)[-1],
\]
which is infinite dimensional for $d>2$. If $d=2$, this space is finite dimensional, so it makes sense to extend the theory down to codimension two.
This is another way of observing that two\ndash dimensional electrodynamics is almost topological (this holds also for nonabelian Yang Mills theories).
\end{Exa}

\subsection{Perturbative quantization}
We may finally present the generalization of the formalism discussed in subsection~\ref{s:pertquant} to the case of degenerate Lagrangians in the BV-BFV
formalism. For simplicity, we assume that the boundary one\ndash form $\alpha^\de_{\de M}$ is globally well-defined and that
$\calF^\de_{\de M}$ is endowed with a Lagrangian foliation on which  $\alpha^\de_{\de M}$  vanishes and which has a smooth leaf space
$\calB_{\de M}$. The space of functions on $\calB_{\de M}$ defines the boundary graded vector space $\calH_{\de M}$.
Let $p_{\de M}$ be the projection $\calF^\de_{\de M}\to \calB_{\de M}$. To produce a state $\psi_M$ associated to the bulk $M$
we first have to choose an embedding of $\underline{\calEL}_{[p_{\de M}^{-1}(\phi)\cap\calL_M]}$ into $\pi_M^{-1}(p_{\de M}^{-1}(\phi))$
and a tubular neighborhood thereof.
Then we have to pick a Lagrangian submanifold $\calL_\phi$ in the  fiber of this tubular neighborhood.
Finally,
\[
\psi_M(\phi)= \int_{\calL_\phi} \EE^{\frac\ii\hbar S_M(\Phi)}\;[D\Phi].
\]
Notice that $\psi_M(\phi)$ is also a function on the moduli space of vacua $\underline{\calEL}_{[p_{\de M}^{-1}(\phi)\cap\calL_M]}$.
As already observed, each of these spaces carries a symplectic structure of degree $-1$ and is by assumption finite dimensional.
The integral has to be computed perturbatively. One can then define a BV operator $\Delta$ on the moduli spaces of vacua and (if the theory is not anomalous)
a coboundary operator $\Omega$ on $\calH_{\de M}$. By general BV arguments we expect that  $\psi_M$ satisfies the following generalization
of the QME
\begin{equation}\label{e:modQME}
\hbar^2\Delta\psi_M + \Omega\psi_M = 0,
\end{equation}
whose classical limit should correspond to \eqref{e:modCME}.

An example where this kind of quantization has been performed and a solution to \eqref{e:modQME} has been explicitly obtained is
described in \cite{AM}. Other examples are currently being studied \cite{CMR-prog}.


\appendix
\section{Some useful facts}\label{a:rec}
A relation from a set $X$ to a set $Y$ is just a subset of $X\times Y$. If $R_1$ is a relation from $X$ to $Y$ and $R_2$ is a relation from $Y$ to $Z$,
the composition $R_2\circ R_1$ from $X$ to $Z$ is defined as
\[
R_2\circ R_1=\{(x,z)\in X\times Z : \exists y\in Y\ (x,y)\in R_1,\ (y,z)\in R_2\}.
\]
The composition is associative. If $\phi\colon X\to Y$ and $\psi\colon Y\to Z$ are maps, then $\graph(\psi)\circ\graph(\phi)=\graph(\psi\circ\phi)$.

If $X$ and $Y$ are symplectic manifolds, a relation from $X$ to $Y$ is called canonical if it is a Lagrangian submanifold. A map $\phi\colon X\to Y$ is a symplectomorphism if{f}
$\graph(\phi)$ is a canonical relation.
The composition of of two canonical relations in general is not a submanifold. On the other hand, being Lagrangian is preserved if $X$ and $Y$ are finite dimensional; otherwise one can only ensure being isotropic in general. A composition of isotropic relations is again isotropic.

In this paper, we often work with presymplectic and weakly symplectic forms. Recall that a closed two-form $\omega$ is presymplectic
if it has constant rank and is weakly symplectic if it defines an injective linear map from the tangent to the cotangent bundle (in finitely many dimensions, this implies that the form is also symplectic).

The notion of  Lagrangian submanifold naturally extends to presymplectic and weakly symplectic manifolds.
A submanifold $L$ of $(M,\omega)$ is called Lagrangian, if $T_xL^\perp=T_xL$ $\forall x\in L$. Here
\[
T_xL^\perp:=\{v\in T_xM\colon \omega_x(v,w)=0\ \forall w\in\T_xL\},
\]
which makes sense also if $\omega_x$ is degenerate.
Similarly, $L\subset M$ is coisotropic if $TL^\perp\subset TL$
and it is isotropic when $TL\subset TL^\perp$. Here $\perp$ means
orthogonal subbundle with respect to the two-form $\omega$.


\begin{thebibliography}{99}
\bibitem{AM} A.~Alekseev and P.~Mn\"ev,
``One-dimensional Chern--Simons theory,'' 
\cmp{307}, 185\Ndash227 (2011).
\bibitem{AKSZ} M. Alexandrov, M. Kontsevich, A. Schwarz and
O. Zaboronsky,
``The Geometry of the Master Equation and Topological Quantum Field Theory,''
\ijmp{A 12}, 1405--1430 (1997).
\bibitem{A} M.~ Atiyah, (1988), ``Topological quantum field theories," Publications Math\'ematiques de l'IHES 68, 175\Ndash186 (1988).
\bibitem{BFV} I. A. Batalin and G. A. Vilkovisky, ``Relativistic
S-Matrix of Dynamical Systems with Boson and Fermion Constraints,''
\pl{69 B}, 309--312 (1977);\\
E. S. Fradkin and T. E. Fradkina, ``Quantization of Relativistic
Systems with Boson and Fermion First- and Second-Class Constraints,''
\pl{72 B}, 343--348 (1978);\\
I. A. Batalin and E. S. Fradkin,
``A Generalized Canonical Formalism and Quantization of Reducible
Gauge Theories,'' \pl{B 122}, 157--164 (1983).
\bibitem{BV} I. A. Batalin, G. A. Vilkovisky, ``Gauge algebra and quantization," Phys.\ Lett.\ \textbf{B 102}, 1 (1981)
27\Ndash31;
I. A. Batalin, G. A. Vilkovisky, ``Quantization of gauge theories with linearly dependent generators," Phys.\ Rev.\ \textbf{D 28}, 10 (1983) 2567\Ndash2582.
\bibitem{Bor} M. Bordemann, ``The deformation quantization of certain super-Poisson brackets and BRST cohomology," arXiv:math.QA/0003218
\bibitem{CMR} A.~S.~Cattaneo, P.~Mn\"ev and N.~Reshetikhin,
``Classical BV theories on manifolds with boundaries,'' 
\href{http://arxiv.org/abs/1201.0290}{math-ph/1201.0290}
\bibitem{CMR-prog} A.~S.~Cattaneo, P.~Mn\"ev and N.~Reshetikhin,
``Perturbative BV-BFV theories on manifolds with boundary," in preparation.
\bibitem{C} D. Christodoulou, \emph{The Action Principle and Partial Differential Equations,} Princeton University Press, 1999.
\bibitem{F} V. Fock, private communication.
\bibitem{FK} J. Fr\"ohlich and C. King, ``The Chern--Simons Theory and
Knot Polynomials,'' \cmp{126}, 167--199 (1989).
\bibitem{Fr} D. Freed, `` Classical Chern-Simons theory, Part 1",
Adv.Math. 113,  237-303(1995).
\bibitem{Gawedzki} K. Gawedzki, ``Classical origin of quantum group symmetries in 
Wess--Zumino--Witten conformal field theory," \cmp{139}, 201\Ndash213 (1991).
\bibitem{G} K. Gawedzki, ``Conformal field theory: A case study," arXiv:hep-th/9904145
\bibitem{Go} M. J. Gotay, ``On coisotropic imbeddings of presymplectic manifolds,'' Proc.\ AMS 84(1), 111\Ndash114 (1982).
\bibitem{HT} M. Henneaux and C. Teitelboim,
\emph{Quantization of Gauge Systems,} PUP, 1992.
\bibitem{Her} H.-C. Herbig, \emph{Variations on homological Reduction,} Ph.D.\ Thesis (University of Frankfurt), arXiv:0708.3598
\bibitem{L} J.~Lurie, ``On the classification of topological field theories,''
arXiv:0905.0465
\bibitem{Mig} A. Migdal, ``Recursion Relations in Gauge Theories," Zh.\ Eksp.\ Teor.\ Fiz. \textbf{69} (1975)
810 (Sov. Phys. Jetp. \textbf{42} 413).
\bibitem{RT} N. Yu. Reshetikhin and V. G. Turaev, ``Invariants of
3-Manifolds via Link Polynomials and Quantum Groups,''
\inm{103} 547--597 (1991).
\bibitem{R} D. Roytenberg, \textit{AKSZ-BV Formalism and Courant Algebroid-induced Topological Field Theories,}
Lett. Math. Phys. 79 (2007) 143--159.
\bibitem{SchBFV} F.~Sch\"atz, ``BFV-complex and higher homotopy structures,"
\cmp{286}, 441\Ndash480 (2009); F.~Sch\"atz,  \emph{Coisotropic Submanifolds and the BFV-Complex}, Ph.D.~thesis,
University of Zurich, 2009: \href{http://user.math.uzh.ch/cattaneo/schaetz.pdf}{http://user.math.uzh.ch/cattaneo/schaetz.pdf}
\bibitem{Schw} A. Schwarz, ``Geometry of Batalin--Vilkovisky quantization,"
\cmp{155}, 249\Ndash260 (1993).
\bibitem{Schwarz} A. Schwarz, ``Symplectic formalism in conformal field theory," in 
\emph{Quantum Symmetries,} Les Houches LXIV (ed. A. Connes, K. Gawedzki, J. Zinn-Justin), 957\Ndash977 (1995). 
\bibitem{S} G.~Segal, Notes on Quantum Field Theory, \href{http://www.cgtp.duke.edu/ITP99/segal/segal1_n.pdf}{http//www.cgtp.duke.edu/ITP99/segal/segal1\_n.pdf}, 1999;
``The Definition of Conformal Field Theory.'' In: \emph{Topology, Geometry and Quantum Field Theory,} London Mathematical Society Lecture Note Series (No.\ \textbf{308}), Cambridge
University Press, 2004, p.\ 421\Ndash577.
\bibitem{StaBFV} J. Stasheff, ``Homological reduction of constrained Poisson algebras," \jdg{45}, 221\Ndash240 (1997).
\bibitem{PTV} B. To\"en, G. Vezzosi, ``From HAG to DAG: derived moduli stacks'', in: \textit{Axiomatic, enriched and motivic homotopy theory}, NATO Science Series 131 II (2004) 173--216.
\bibitem{W} E.~Witten, ``Quantum Field Theory and the Jones Polynomial,''
\cmp{121}, 351--399 (1989).
\end{thebibliography}
\end{document}